\documentclass[aps,prl,showpacs,twocolumn]{revtex4}
\usepackage{amssymb}
\usepackage{epsfig,graphicx}

\begin{document}

\title{Localized Asymmetric Atomic Matter Waves in Two-Component
Bose-Einstein Condensates Coupled with Two Photon Microwave Field}
\author{Bo Xiong}
\affiliation{Beijing National Laboratory for Condensed Matter Physics, Institute of
Physics, Chinese Academy of Sciences, Beijing 100080, China}
\date{\today}

\begin{abstract}
We investigate localized atomic matter waves in two-component Bose-Einstein
condensates coupled by the two photon microwave field. Interestingly, the
oscillations of localized atomic matter waves will gradually decay and
finally become non-oscillating behavior even if existing coupling field. In
particular, atom numbers occupied in two different hyperfine spin states
will appear asymmetric occupations after some time evolution.
\end{abstract}

\pacs{03.75.-b,67.40.-w,39.25.+k}
\maketitle

\address{Beijing National Laboratory for Condensed Matter Physics,
Institute of Physics, Chinese Academy of Sciences, Beijing 100080,
China}

The creation of dark solitons \cite{ReinhardtJPB,MorganPRA,BuschPRL2001} and
bright solitons \cite{StreckerNature,KhaykovichScience} in trapped weakly
interacting atomic gases has opened the road to understanding and
controlling nonlinear properties of atomic matter waves. Although many
nonlinear phenomena in Bose-Einstein condensates (BECs) may have their
counterparts in nonlinear optics, unique manageability of the properties of
BECs, new setups can be studied, which were not realized in nonlinear
optics. This opens the promising perspective for numerous applications of
the nonlinear matter-wave physics such as atom chip and quantum information
processing on the nanometer scale. The s-wave scattering interaction in the
BECs is the determining factor (attractive for bright solitons, repulsive
for dark solitons) and attractive interaction leads to condensates collapse
in effective two- and three- dimensional cases if the atomic number exceeds
a critical value, and to soliton formation in quasi one-dimensional (1D)
traps. In the quasi-1D regime, two condensates in different hyperfine states
reduce to coupled Gross-Pitaevskii equations (GPEs) \cite%
{BuschPRL2000,ModugnoPRE}. In the system of optically coupled two-component
BECs, some beautiful work \cite{WilliamsPRA,ParkPRL,KohlerPRL} has discussed
Josephson-type density evolution process and pointed out the internal
tunnelling effects of the two optically coupled condensates.

In the present Letter, we consider two-component Bose-Einstein condensates
coupled by two photon microwave field. Using the soliton as staring spatial
pattern formation, we study dynamic behavior of two components condensates
under tuning atomic interactions. Surprisingly, the spatial shape of
condensates presents very localized all the time even if changing the atomic
interactions. During the transmission process, two localized atomic matter
waves will collide each other many times and exchange energy greatly. In
particular, oscillations of atom numbers of two components are very unstable
and will periodically decay to zero, and the speed of decay can also be
controlled through tuning atomic interactions. Moreover, atom numbers
occupied in two different hyperfine spin states will appear asymmetric
occupations in the end.

Experimentally, the bright soliton have been successfully produced from $%
^{7}Li$ atoms in the internal atomic hyperfine spin state $%
|F=1,m_{f}=1\rangle $ \cite{StreckerNature,KhaykovichScience}.
Theoretically, we consider two different hyperfine spin state with
attractive interactions denoted as $\Psi _{1}$ and $\Psi _{2}$,
respectively. The ratio of the radial and the axial trapping frequency of
harmonic potential obeys $\omega _{r}/\omega _{z}\gg 1$, for example, $%
\omega _{r}/\omega _{z}=5.1\times 10^{3}$) \cite{StreckerNature}, so two
hyperfine spin states are both in the transverse ground state. The
condensate wave function is normalized as $\int dr(|\Psi _{1}|^{2}+|\Psi
_{2}|^{2})=N_{1}+N_{2}=N$, where $N$ is the total number of atom in two
components. The attractive intra- and interatomic interactions are $%
U_{ii}=-4\pi \hbar ^{2}|a_{ii}||\Psi _{i}|^{2}/m$ and $U_{ij}=-4\pi \hbar
^{2}|a_{ij}||\Psi _{i}|^{2}/m$ $(i,j=1,2)$, where $-|a_{ii}|$ and $-|a_{ij}|$
is the intra-species and inter-species s-wave scattering lengths, m is the
atomic mass of the $^{7}Li$ atom. In one-dimensional case the effective
couplings $\bar{U}_{ii}$ and $\bar{U}_{ij}$ is represented by $\bar{U}%
_{ii}=-2\hbar ^{2}|a_{ii}||\Psi _{i}|^{2}/ma_{r}^{2}$ and $\bar{U}_{ij}$ $%
=-2\hbar ^{2}|a_{ij}||\Psi _{i}|^{2}/ma_{r}^{2}$ where $a_{r}=\sqrt{\hbar
/m\omega _{r}}$ \cite{OlshaniiPRL}. After tuning the strength of interaction
to a sufficiently large negative value through use of atomic Feshbach
resonance \cite{InouyeScience,CornishPRL}, the condensates can be set free
along the waveguide, i.e. $\omega _{z}=0$. The atomic transition between
two-component BECs is induced by the two photon microwave field with the
effective Rabi frequency $\Omega $ and a finite detune $\delta $, where the $%
\Omega $ and $\delta $ are independent of both time and axial coordinate as
experimental case \cite{MatthewsPRL1998}.

In view of the difference of inter- and intra-atomic interactions, it is
convenient to describe the two-component BECs coupled with the two photon
microwave field in the following dimensionless coupled GPEs,
\begin{eqnarray}
&&i\frac{\partial \phi _{i}}{\partial t}=-\frac{\partial ^{2}\phi _{i}}{%
\partial z^{2}}-\frac{4N\left\vert a_{ij}\right\vert }{a_{r}}(|\phi
_{i}|^{2}+|\phi _{j}|^{2})\phi _{i}-\frac{\delta }{\omega _{r}}\cos (\pi
i)\phi _{i}  \nonumber \\
&&+\frac{\Omega }{\omega _{r}}\phi _{j}-C_{i}\frac{4N\left\vert
a_{ij}\right\vert }{a_{r}}|\phi _{i}|^{2}\phi _{i}.  \label{GP}
\end{eqnarray}%
where $i,j=1,2$, $i\neq j$ and time t and coordinate z are measured in units
$2/\omega _{r}$ and $a_{r}=\sqrt{\hbar /m\omega _{r}}$, respectively, so,
energy is in unit of $\hbar \omega _{r}/2$, $C_{i}=(\left\vert
a_{ii}\right\vert -\left\vert a_{ij}\right\vert )/\left\vert aij\right\vert $
measure the difference of interaction strength, and $\int dz(|\phi
_{1}|^{2}+|\phi _{2}|^{2})=1$ is normalization condition.

In order to compare with experimental case, some parameters in equations (%
\ref{GP}) are chosen as $\omega _{r}=2\pi \times 625$ $Hz$ (so, $%
a_{r}\approx 1.51$ $\mu m$), $\left\vert a_{ij}\right\vert =0.16$ $nm$ (same
as \cite{StreckerNature}). Since $a_{r}/\left\vert a_{ij}\right\vert \approx
9437$ \cite{PerezPRA,Muryshev,Carr,Kivshar}, so we choose the number of $%
^{7}Li$ atom $N=6000$, and so coefficient $4N \vert a_{ij} \vert /a_{r}
\approx 2.5$.

In the special case of $C_{i}=0$, i.e. $|a_{ii}|=|a_{ij}|$, equations (\ref%
{GP}) have exactly two-soliton solutions
\begin{eqnarray}
\phi _{i} &=&[\sin (\frac{\pi i}{2}-\theta )e^{i\Gamma t}(\alpha _{i}e^{\eta
_{i}}+\alpha _{j}e^{\eta _{j}}+e^{\delta _{i}+\eta _{i}+\eta _{i}^{\ast
}+\eta _{j}}  \nonumber \\
&&+e^{\delta _{j}+\eta _{i}+\eta _{j}+\eta _{j}^{\ast }})-\cos (\frac{\pi i}{%
2}-\theta )e^{-i\Gamma t}(\beta _{i}e^{\eta _{i}}+\beta _{j}e^{\eta _{j}}
\nonumber \\
&&+e^{\delta _{i}+\eta _{i}+\eta _{i}^{\ast }+\eta _{j}}+e^{\delta _{j}+\eta
_{i}+\eta _{j}+\eta _{j}^{\ast }})]/D.  \label{Soliton2}
\end{eqnarray}%
here $i,j=1,2$, $i\neq j$ and $\theta =\tan ^{-1}(\frac{\Omega }{\delta })/2$%
, $\Gamma =[(\delta /\omega _{r})^{2}+(\Omega /\omega _{r})^{2}]^{1/2}$, $%
\eta _{j}=k_{j}(z+ik_{j}t)$, $D=1+e^{\eta _{1}+\eta _{1}^{\ast
}+R_{1}}+e^{\eta _{1}+\eta _{2}^{\ast }+\delta _{0}}+e^{\eta _{1}^{\ast
}+\eta _{2}+\delta _{0}^{\ast }}+e^{\eta _{2}+\eta _{2}^{\ast
}+R_{2}}+e^{\eta _{1}+\eta _{1}^{\ast }+\eta _{2}+\eta _{2}^{\ast }+R_{3}}$,
$e^{\delta _{0}}=\kappa _{12}/(k_{1}+k_{2}^{\ast })$, $e^{R_{j}}=\kappa
_{jj}/(k_{j}+k_{j}^{\ast })$, $e^{\delta _{i}}=(k_{i}-k_{j})(\alpha
_{i}\kappa _{ji}-\alpha _{j}\kappa _{ii})/(k_{i}+k_{i}^{\ast })(k_{i}^{\ast
}+k_{j})$, $e^{\delta _{i}^{\prime }}=(k_{i}-k_{j})(\beta _{i}\kappa
_{ji}-\beta _{j}\kappa _{ii})/(k_{i}+k_{i}^{\ast })(k_{i}^{\ast }+k_{j})$, $%
e^{R_{3}}=|k_{1}-k_{2}|^{2}(\kappa _{11}\kappa _{22}-\kappa _{12}\kappa
_{21})/(k_{1}+k_{1}^{\ast })(k_{2}+k_{2}^{\ast })|k_{1}+k_{2}^{\ast }|^{2}$,
and $\kappa _{ij}=[(2N\left\vert a_{12}\right\vert /a_{r})(\alpha _{i}\alpha
_{j}^{\ast }+\beta _{i}\beta _{j}^{\ast })][(k_{i}+k_{j}^{\ast })]$.$\left[
(\alpha _{1},\beta _{1})\{\kappa _{11}^{-1},\kappa _{12}^{-1}\}-(\alpha
_{2},\beta _{2})\{\kappa _{21}^{-1},\kappa _{22}^{-1}\}\right] $

In order to facilitate the understanding of the influences of microwave
field on solitons, it is convenient to obtain the asymptotic forms of
solutions (\ref{Soliton2}) of Eq. (\ref{GP})
\begin{equation}
S_{j}^{n\pm }=[\sin (\frac{\pi j}{2}-\theta )e^{i\Gamma t}A_{j}^{n\pm }-{%
\cos (}\frac{\pi j}{2}-\theta )e^{-i\Gamma t}A_{j}^{n\pm }]\phi ^{n\pm }.
\label{Soliton2_2}
\end{equation}%
here superscripts in $S_{j}^{n\pm }$ $(j,n=1,2)$ denote solitons in limit $%
t\rightarrow \pm \infty $, subscripts refer to the components and $(\phi
^{1-}$,$\phi ^{2+})=\{k_{1R}e^{i\eta _{1I}}\text{sech}(\eta _{1R}+R_{1}/2)$,
$k_{2R}e^{i\eta _{2I}}$sech$(\eta _{2R}+R_{2}/2)\}$, $(\phi ^{2-}$,$\phi
^{1+})$ $=\{k_{2R}e^{i\eta _{2I}}\text{sech}[\eta _{2R}+(R_{3}-R_{1})/2]$, $%
k_{1R}e^{i\eta _{1I}}$sech$[(\eta _{1R}+(R_{3}-R_{1})/2]\}$, and the
polarization vectors $A_{i}^{j-}$ and $A_{i}^{j+}$are defined as $%
\{(A_{1}^{1-},A_{2}^{1-}),(A_{1}^{2+},A_{2}^{2+})\}=[(2N\left\vert
a_{12}\right\vert /a_{r})(|\alpha _{1}|^{2}+|\beta _{1}|^{2})]^{\frac{1}{2}%
}\{(\alpha _{1},\beta _{1}),(\alpha _{2},\beta _{2})\}$ and $%
\{(A_{1}^{2-},A_{2}^{2-}),(A_{1}^{1+},A_{2}^{1+})\}=c[(2N\left\vert
a_{12}\right\vert /a_{r})(|\alpha _{1}|^{2}+|\beta _{1}|^{2})^{\frac{1}{2}%
}]\{(a_{1}/a_{1}^{\ast }),(a_{2}/a_{2}^{\ast })\}\{(\alpha _{1}\kappa
_{11}^{-1}-\alpha _{2}\kappa _{21}^{-1},\beta _{1}\kappa _{11}^{-1}-\beta
_{2}\kappa _{21}^{-1}),(\alpha _{1}\kappa _{12}^{-1}-\alpha _{2}\kappa
_{22}^{-1},\beta _{1}\kappa _{12}^{-1}-\beta _{2}\kappa _{22}^{-1})\}$ in
which $c=(1/|\kappa _{12}|^{2}-1/\kappa _{11}\kappa _{22})^{-\frac{1}{2}}$, $%
a_{i}=(k_{i}+k_{j}^{\ast })\left[ \cos (\pi i+\pi )(k_{i}-k_{j})(\alpha
_{i}^{\ast }\alpha _{j}+\beta _{i}^{\ast }\beta _{j})\right] ^{\frac{1}{2}}$%
.
\begin{figure}[tbp]
\epsfxsize=9.5cm \centerline{\epsffile{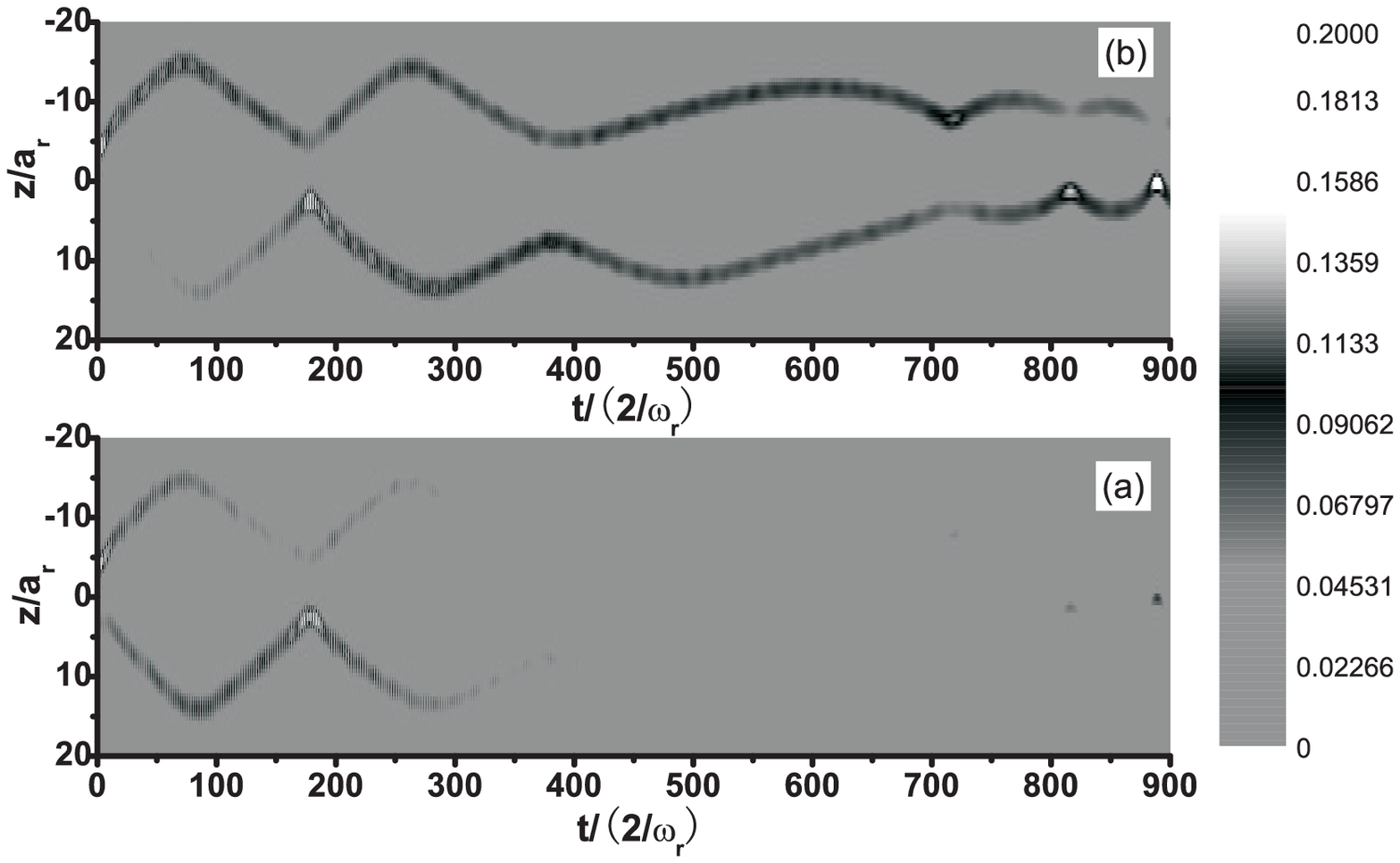}}
\caption{Time development of the condensate densities $(a)$ $%
|\protect\phi _{1}|^{2}$, $(b)$ $|\protect\phi _{2}|^{2}$ along the $z$ axis
for $N=6000$, $\protect\omega _{r}=2\protect\pi \times 625$ $Hz$, $%
a_{r}\approx 1.51$ $\protect\mu m$, {\protect\LARGE \ }$\protect\delta =0.25%
\protect\omega _{r}${\protect\LARGE \ }and{\protect\LARGE \ }$\Omega =0.75%
\protect\omega _{r}$, $a_{11}=-0.24$ $nm$, $a_{12}=-0.16$ $nm$, $%
a_{22}=-0.32 $ $nm$. The chosen parameters for initial solitons are $%
k_{1}=0.313+0.2i$, $k_{2}=0.313-0.2i$, $\protect\alpha _{1}=\protect\beta %
_{1}=\protect\beta _{2}=0.251+0.966i$, $\protect\alpha _{2}=0.966+0.251i$.}
\end{figure}

\begin{figure}[tbp]
\epsfxsize=9.5cm \centerline{\epsffile{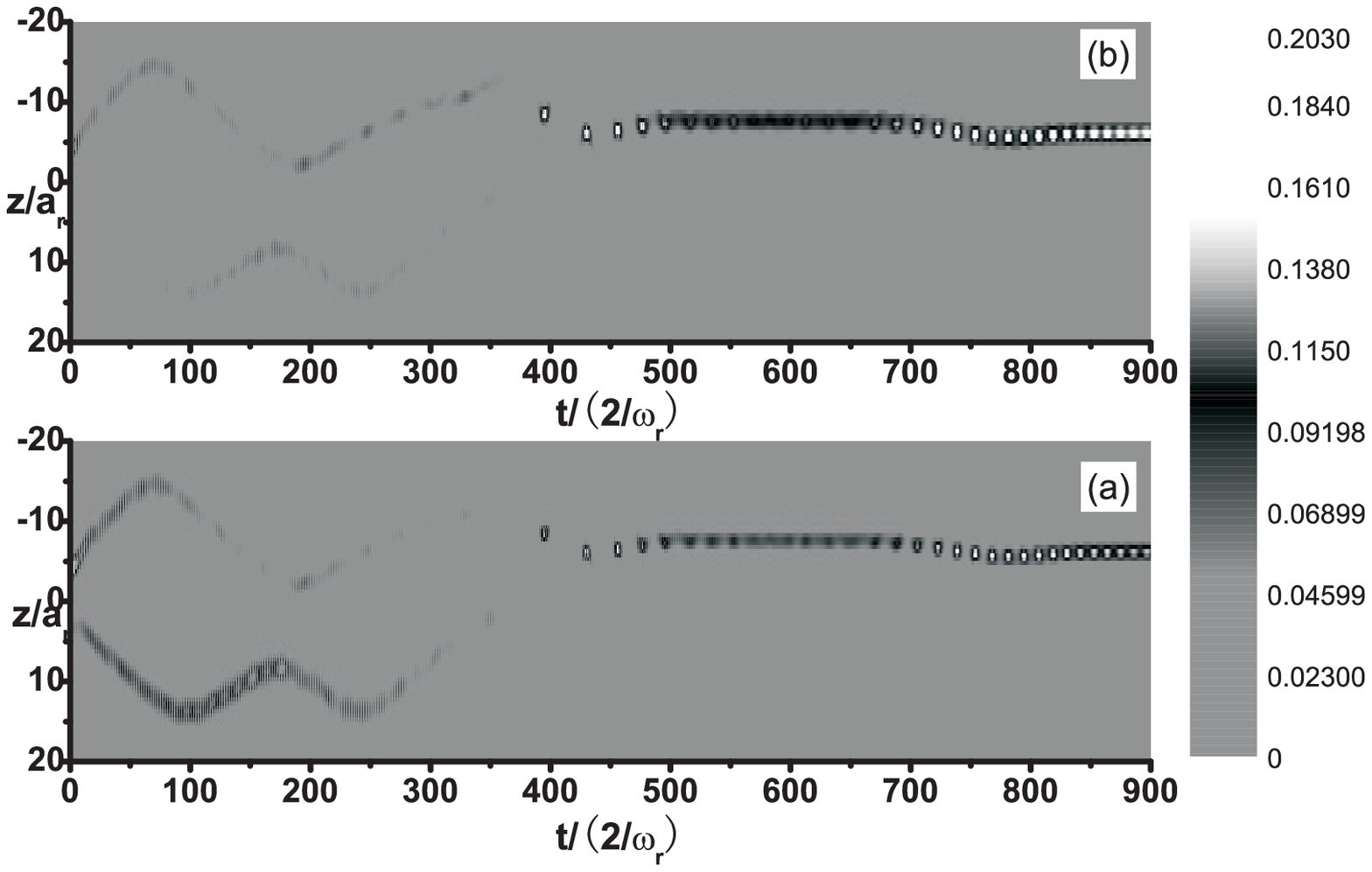}}
\caption{Time development of the condensate densities $(a)$ $%
|\protect\phi _{1}|^{2}$, $(b)$ $|\protect\phi _{2}|^{2}$ along the $z$ axis
for $N=6000$, $\protect\omega _{r}=2\protect\pi \times 625$ $Hz$, $%
a_{r}\approx 1.51$ $\protect\mu m$, {\protect\LARGE \ }$\protect\delta =0.25%
\protect\omega _{r}${\protect\LARGE \ }and{\protect\LARGE \ }$\Omega =0.75%
\protect\omega _{r}$, $a_{11}=-0.32$ $nm$, $a_{12}=-0.16$ $nm$, $%
a_{22}=-0.24 $ $nm$. The chosen parameters for initial solitons are $%
k_{1}=0.313+0.2i$, $k_{2}=0.313-0.2i$, $\protect\alpha _{1}=\protect\beta %
_{1}=\protect\beta _{2}=0.251+0.966i$, $\protect\alpha _{2}=0.966+0.251i$.}
\end{figure}

It is easy to see from the Eqs. (\ref{Soliton2_2}) that the density of
solitons are $\left\vert S_{l}^{n\mp }(z,t)\right\vert ^{2}=[|A_{l}^{n\mp
}|^{2}\cos ^{2}\theta +|A_{j}^{n\mp }|^{2}\sin ^{2}\theta
+(-1)^{l}|A_{l}^{n\mp }||A_{j}^{n\mp }|\sin 2\theta \cos (2\Gamma t+\varphi
^{n\mp })]\left\vert \phi ^{n\mp }\right\vert ^{2}$,$\;n,l,j=1,2,$ $l\neq j$%
, where $\varphi ^{n\mp }=\tan ^{-1}\left( A_{1I}^{n\mp }/A_{1R}^{n\mp
}\right) -\tan ^{-1}\left( A_{2I}^{n\mp }/A_{2R}^{n\mp }\right) $. So, in
this special case, the reason of density oscillation is just because of the
presence of time dependence part $(-1)^{l}|A_{l}^{n\mp }||A_{j}^{n\mp }|\sin
2\theta \cos (2\Gamma t+\varphi ^{n\mp })\phi ^{n\mp }(z,t)^{2}$, where the
frequency of oscillation is $2\Gamma $ and can be tuned easily through
changing the effective Rabi frequency $\Omega $, a finite detune $\delta $
and radial trapping frequency $\omega _{r}$. In particular, when $\theta
=n\pi /2$ $(n\in even)$ or $\Gamma =0$, the oscillation will disappear.

More interesting case is see the dynamics caused by changing the
nonlinearities, i.e. atomic interactions. Experimentally, changing the
nonlinearity can be realized by Feshbach resonance. In this system, changing
the nonlinearity corresponds to change $C_{i}$ in equations (\ref{GP}). So,
we will investigate pattern formation under changing $C_{i}$. We perform
numerical simulations of Eq. (\ref{GP}) based on discretizing it in both
space and time using the known boundary conditions, e.g., $\left\vert \phi
_{j}(z_{\min },t)\right\vert =\left\vert \phi _{j}(z_{\max },t)\right\vert
=0 $ where $z_{\min }=-20$ and $z_{\max }=20$. Initial states we prepared
are two soliton solutions (\ref{Soliton2}) of Eq. (\ref{GP}). The mismatch
of $a_{ii}$ with $a_{ij}$ in initial configuration will result in
perturbations of the condensates. In our numerical simulations, the
parameters are chosen as $k_{1}=0.313+0.2i$, $k_{2}=0.313-0.2i$, $\alpha
_{1}=\beta _{1}=\beta _{2}=0.251+0.966i$, $\alpha _{2}=0.966+0.251i$, $%
\delta =0.25 \Omega _{r}$ and $\Omega =0.75\omega _{r}$. So, the frequency
of oscillations is $2\Gamma =2[(\delta /\omega _{r})^{2} +(\Omega /\omega
_{r})^{2}]^{1/2}=1.58$ corresponding to $0.8$ $msec$ considering time units.

Figure 1 shows image of initial conditions (\ref{Soliton2}) propagating
about $560$ periods for $\left\vert a_{11}\right\vert =1.5\left\vert
a_{12}\right\vert =0.24$ $nm$ and $\left\vert a_{22}\right\vert =2\left\vert
a_{12}\right\vert =0.32$ $nm$. It is interesting to point out that in our
numerical simulations, spatial pattern formation all the time preserves
soliton-like shaped i.e., localized atomic matter waves during transmission
process. Although introduced important the mismatch of $a_{ii}$ with $a_{ij}$%
, their influences on dynamics of initial solitons are only velocity and
amplitude. That is to say, increasing the absolute value of atomic
interactions will make the condensates more local instead of destroying
localized density distributions.

As is shown in Figure1, in the beginning, the densities of two localized
atomic matter waves are oscillate periodically and later, oscillations will
quickly damp out for each components as shown in Fig. 1(a) and Fig. 1(b),
respectively. The total density of two components is well conserved during
the transmission process, but the densities of each components will change
significantly. This is very incompatible with special case $\vert a_{11}
\vert = \vert a_{12} \vert =\vert a_{22} \vert$.

One of the most different phenomenon from special case is that two localized
waves for each component no longer transmit separately and on the contrary,
they transmit closer and closer. We can see that the two localized waves
collide for the first time at about $90$ $msec$. After the collision, the
relative density of one of waves is enhanced and the other is suppressed
obviously. As shown in Figure 1, the density of component one corresponding
to Fig. 1(a) is suppressed and the other corresponding to Fig. 1(b) is
enhanced at the same time. So, the interaction between two waves can be used
to transfer energy from one to the other. At about $190$ $msec$, the two
waves collide for the second time and oscillations are very weak now.
Subsequently, the two waves will keep to collide faster and become closer
and closer.

It must be pointed out that whether the atomic interactions have the same
values or not, the two localized atomic matter waves for component one are
always completely overlapped with the others for component two in
configuration space. So, from the experimental point of view, observable
phenomena are always the combination effects of two components. The
oscillations effect as shown in Fig. 1(a) and Fig. 1(b) can be observed
through kicking BECs so that different hyperfine spin components will be
separated in configuration space.

In two different hyperfine spin components, two localized atomic matter
waves are move closer and closer after a few collisions. So, now the
questions are can we make two waves close faster and transfer energy more
efficient? In Figure 1, the atomic s-wave scattering lengths are chosen as $%
a_{11}=-0.24$ $nm$, $a_{12}=-0.16$ $nm$, $a_{22}=-0.32$ $nm$ and the
relation of absolute value of $a_{11}$ and $a_{22}$ is $\left\vert
a_{11}\right\vert <\left\vert a_{22}\right\vert $. So, it is naturally to
investigate the time development of condensates in condition of inversing
the relation of $\left\vert a_{11}\right\vert $ and $\left\vert
a_{22}\right\vert $, i.e. $\left\vert a_{11}\right\vert >\left\vert
a_{22}\right\vert $. It is very surprising that two localized waves not only
close faster but also transfer energy more efficient than former case. This
is illustrated in Figure 2. The atomic s-wave scattering lengths are chosen
as $a_{11}=-0.32$ $nm$, $a_{12}=-0.16$ $nm$, $a_{22}=-0.24$ $nm$, and the
other parameters are chose as the same as before.

\begin{figure}[tbp]
\epsfxsize=9.5cm \centerline{\epsffile{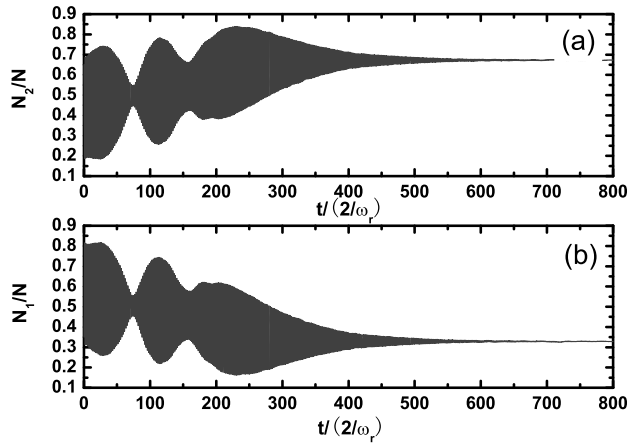}} \caption{Time
development of the atom numbers $(a)$
component $\protect\phi _{1}$, $(b)$ component $\protect\phi _{2}$ for $%
N=6000$, $\protect\omega _{r}=2\protect\pi \times 625$ $Hz$, $a_{r}\approx
1.51$ $\protect\mu m$, {\protect\LARGE \ }$\protect\delta =0.25\protect%
\omega _{r}${\protect\LARGE \ }and{\protect\LARGE \ }$\Omega =0.75\protect%
\omega _{r}$, $a_{11}=-0.24$ $nm$, $a_{12}=-0.16$ $nm$, $a_{22}=-0.32$ $nm$.
The chosen parameters for initial solitons are $k_{1}=0.313+0.2i$, $%
k_{2}=0.313-0.2i$, $\protect\alpha _{1}=\protect\beta _{1}=\protect\beta %
_{2}=0.251+0.966i$, $\protect\alpha _{2}=0.966+0.251i$.}
\end{figure}

\begin{figure}[tbp]
\epsfxsize=9.5cm \centerline{\epsffile{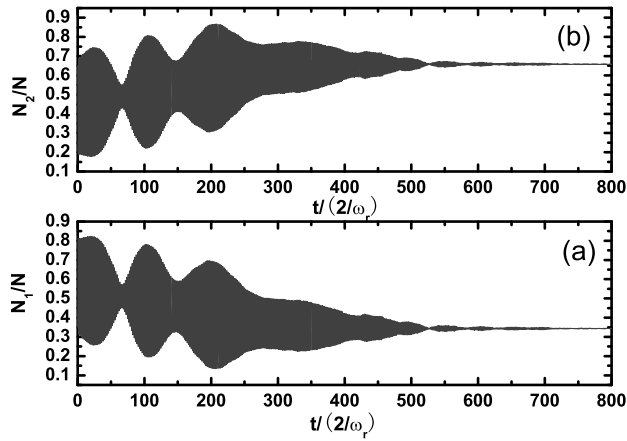}} \caption{Time
development of the atom numbers $(a)$
component $\protect\phi _{1}$, $(b)$ component $\protect\phi _{2}$ for $%
N=6000$, $\protect\omega _{r}=2\protect\pi \times 625$ $Hz$, $a_{r}\approx
1.51$ $\protect\mu m$, {\protect\LARGE \ }$\protect\delta =0.25\protect%
\omega _{r}${\protect\LARGE \ }and{\protect\LARGE \ }$\Omega =0.75\protect%
\omega _{r}$, $a_{11}=-0.32$ $nm$, $a_{12}=-0.16$ $nm$, $a_{22}=-0.24$ $nm$.
The chosen parameters for initial solitons are $k_{1}=0.313+0.2i$, $%
k_{2}=0.313-0.2i$, $\protect\alpha _{1}=\protect\beta _{1}= \protect\beta %
_{2}= 0.251+0.966i$, $\protect\alpha _{2}=0.966+0.251i$.}
\end{figure}

Fig. 2(a) and Fig. 2(b) show the time development of the condensate
densities $|\phi _{1}|^{2}$ and $|\phi_{2}|^{2}$ along the $z$ axis,
respectively. Due to changing the relation from $\vert a_{11} \vert < \vert
a_{22} \vert$ to $\vert a_{11} \vert > \vert a_{22} \vert$, the density of
one of localized waves has enhanced greatly and the other has suppressed
extremely only after the second time collision. And the density of enhanced
localized wave of each component will no longer increase or decrease during
transmission process after the second time collision. Moveover, enhanced
localized wave of each component will fixed in the position all the time and
the weaken localized wave will move towards the boundary after the second
time collision. The future applications of these effects can be existed in
atom optics, atom transport and quantum information.

Another interesting problem is the effects of changing atomic interactions
on atom numbers of each components. From Eq. (\ref{Soliton2_2}), it is
easier to see that due to applied coupling field, atom numbers of each
hyperfine component will exchange periodically. Naturally, the question is
whether the persistent oscillations can still be observed or not when tuning
intra- and inter- atomic interactions. Figure 3 and 4 show the time
development of atom numbers of each component for different atomic
interactions. In Figure 3, as the system evolves, oscillations of each
component both enhanced gradually in the beginning. At about $10$ $msec$,
the intensities of oscillations of each component both achieve maximum, and
then begin to decrease, until at about $38$ $msec$ achieve minimum. At time
interval of $38$ $msec$ to $80$ $msec$, the intensities of oscillations
again change from maximum to minimum. Then, it is very surprising that after
achieving maximum for the third time at about $115$ $msec$, oscillations
will gradually damp out and disappear finally. Moreover, it is interesting
to notice that more atom number occupies the hyperfine spin state $\phi _{2}$
as shown in Fig. 3(b) and less atom number occupies the other hyperfine spin
state $\phi _{1}$ as shown in Fig. 3(a) when oscillations disappear. The
ratio of atom numbers of each component is $0.32:0.68$.

The same effects can be observed in Figure 4. When the relation of atomic
interactions changes as shown in Figure 4, oscillations will experience more
periods than former case before damp out. The ratio of atom numbers of each
component is $0.35: 0.65$ and slightly different from former case.

So, the conclusion is that atomic interactions have strong effects on
exchange of atom numbers in this system. Through tuning atomic interactions
using Feshbach resonance, we can control relative atom numbers occupied
different hyperfine spin states and their oscillations. That is to say, we
can control total magnetic moment of condensates using above mentioned
effects. It is very fascinating to observe these effects experimentally.

In this Letter, we have found a new type of wave -- localized atomic matter
waves in two-component BECs coupled with the two photon microwave field. By
tuning the atomic interactions, localized atomic matter waves will
experience a damped oscillations. During the transmission process, two
localized atomic matter waves will collide many times and exchange energy
significantly. In particular, the atom numbers occupied different hyperfine
spin states will appear asymmetric transition finally. Our result gives some
fascinating effects on dynamics which are unobserved in real experiments so
far. This work may be important for controllable nonlinear atom optics.

This work was supported by the NSF of China under grant 90403034, 90406017,
60525417; and the National Key Basic Research Special Foundation of China
under 2005CB724508 and 2006CB921400.

\end{document}